\documentclass[10pt, twoside]{article}

\usepackage[tbtags]{amsmath}
\usepackage{accents}
\usepackage{rotating}
\usepackage{amsfonts}
\usepackage{amssymb}
\usepackage{latexsym}

\DeclareMathSymbol{\vecarrow}{\mathord}{letters}{"7E}

\newlength{\lvech}
\newlength{\lvecw}

\newcommand{\lvec}[1]{\ensuremath{
\text{\settoheight{\lvech}{$\vecarrow$}\addtolength{\lvech}{-.047ex}\settowidth{\lvecw}{$\vecarrow$}$\accentset{\hspace{.47\lvecw}\begin{rotate}{180}\makebox[0pt]{\raisebox{-\lvech}[0pt][0pt]{$\vecarrow$}}\end{rotate}}{#1}$}}}

\newcommand{\starMP}{*_{\scriptscriptstyle MC}}
\newcommand{\starM}{*_{\scriptscriptstyle M}}
\newcommand{\starP}{*_{\scriptscriptstyle C}}

\newcommand{\fett}[1]{\mbox{\boldmath$#1$}} 

\newcommand{\beq}{\begin{equation}}
\newcommand{\eeq}{\end{equation}}

\begin{document}
\makeatletter
\title{Star Products and Geometric Algebra}
\author{
Peter Henselder\footnote{henselde@dilbert.physik.uni-dortmund.de}
,\,\,
Allen C.\ Hirshfeld\footnote{hirsh@physik.uni-dortmund.de}\, and
Thomas Spernat\footnote{tspernat@zylon.physik.uni-dortmund.de} \\
Fachbereich Physik, Universit\"at Dortmund\\
44221 Dortmund}

\maketitle

\begin{abstract}
The formalism of geometric algebra can be
described as deformed super analysis. The deformation is done with
a fermionic star product, that arises from deformation
quantization of pseudoclassical mechanics. If one then extends the
deformation to the bosonic coefficient part of superanalysis one
obtains quantum mechanics for systems with spin. This approach
clarifies on the one hand the relation between Grassmann and
Clifford structures in geometric algebra and on the other hand the
relation between classical mechanics and quantum mechanics.
Moreover it gives a formalism that allows to
handle classical and quantum mechanics in a consistent manner.
\end{abstract}

\section{Introduction}
\qquad Geometric algebra goes back to early ideas of Hamilton,
Grassmann and Clifford. But it was first developed into a full
formalism by Hestenes in \cite{Hestenes1} and \cite{Hestenes3}.
The formalism of geometric algebra is based on the definition of
the geometric or Clifford product. This product is for vectors
defined as the sum of the scalar and the wedge product and 
equips the vector space
with the algebraic structure of a Clifford algebra. This
structure then proved to be a very powerful tool, that allows to
describe and generalize the structures of vector analysis, of
complex analysis and of the theory of spin in a unified and clear
formalism. The formalism can then be used to describe classical
mechanics in the realm of geometric algebra instead of linear
algebra, which is advantageous in many respects
\cite{Hestenes2,Doran1}. It is also possible to generalize the formalism
from the algebra of space to the algebra of spacetime in order to 
describe electrodynamics and special relativity 
\cite{Hestenes1,Doran1}.

In quantum mechanics the Clifford structures of the $\sigma$- and
the $\gamma$-matrices correspond to the structures of
geometric algebra. So by formulating classical physics and quantum
physics with geometric algebra one achieves a formal unification
of both areas on a geometric level. Nevertheless this formulation
is conceptually not totally unified, because classical mechanics
is still formulated on the phase space while quantum mechanics is
formulated in Hilbert space. In order to achieve a totally unified
formulation we will here combine geometric algebra with the star
product formalism. The star product formalism \cite{Bayen} appears
in the context of deformation quantization where one describes the
non-commutativity that enters physics in quantum mechanics not by
using non-commuting objects like operators, but by introducing a
non-commutative product on the phase space that replaces the
conventional product of functions. This star product is so constructed
that the quantized star product of two phase space
functions corresponds to the operator product of the quantized
factors, which then allows to do quantum mechanics
on phase space. To include spin in this formalism we used in 
\cite{Deform3} and \cite{Deform5} fermionic star products that 
result directly from deformation quantization of pseudoclassical
mechanics \cite{Berezin1}. Fermionic star products
were already discussed in \cite{Bayen}, where it was noticed that
they lead to a cliffordization of the underlying Grassmann
algebra. So it is possible to describe a Clifford algebra as a
deformed Grassmann algebra, where this deformation is nothing else
than Chevalley cliffordization \cite{Deform5}.

In this paper we will use the fact that geometric algebra can
be formulated in terms of a Grassmann algebra \cite{Doran2}. We will
show that in this context the geometric product can be made 
explicit as a fermionic star product. It is then straight forward to
translate classical mechanics described with geometric algebra
into a version where it is described in terms of fermionic deformed super
analysis. The fermionic part of the formalism represents hereby the 
basis vector structure of the space on which the theory is formulated, 
i.e.\ the three dimensional space, the phase space or the spacetime. 
In all cases we consider only the case of flat spaces. 
In a second step one can then go over to quantum mechanics,
where we use here deformation quantization, while in \cite{Doran2} 
canonical and path integral quantization was used. Combining in this
way geometric algebra formulated with a fermionic star product with the
bosonic star product of deformation quantization one arrives at a 
supersymmetric star product formalism that allows to describe quantum 
mechanics with spin in a unified manner. Moreover by using star 
products one can immediately give the classical $\hbar\rightarrow 0$ limit 
and see how the spin as a physical observable vanishes. Furthermore
one can see that classical mechanics can be described as a half
deformed theory, while quantum mechanics is a totally deformed theory, 
i.e.\ in classical mechanics the star product acts only on the 
fermionic basis vector part of the formalism, while for $\hbar>0$
there exists also a bosonic star product that acts on the
coefficients of the basis vectors.
  
In the second section we will very shortly review the bosonic and fermionic
star product formalism and show how quantum mechanics with spin can be
described in this context. Then we will show how geometric algebra can
be formulated with the fermionic star product. We will therefore formulate
well known results of geometric algebra in the formalism of fermionic
deformed superanalysis. Afterwards in 
section 5 and 6 we will extend the formalism to the case of 
nonrelativistic quantum mechanics and Dirac theory by using the bosonic
Moyal product.

\section{The Star Product Formalism}
\setcounter{equation}{0}\label{BFosc}
\qquad We first want to introduce the star product
formalism in bosonic and fermionic physics with the example of the
harmonic oscillator \cite{Bayen}. The bosonic oscillator with the 
Hamilton function $H(q,p)=\frac{p^2}{2m}+\frac{m\omega^2}{2}q^2$, can 
be quantized by using the Moyal product
\begin{equation}
f\starM g = f\exp\left[\frac{i\hbar}{2}
\left(\lvec{\partial}_{q}\vec{\partial}_{p}-
\lvec{\partial}_{p}\vec{\partial}_{q}\right)\right]g.
\label{starMDef}
\end{equation}
The star product replaces the conventional product between
functions on the phase space and it is so constructed that the
star anticommutator, i.e.\ the antisymmetric part of first order, is
the Poisson bracket:
\begin{equation}
\lim_{\hbar\rightarrow 0}\frac{1}{i\hbar}
\left[f(q,p),g(q,p)\right]_{\starM}=\lim_{\hbar\rightarrow 0}
\frac{1}{i\hbar}\left(f(q,p)\starM g(q,p)-g(q,p)\starM
f(q,p)\right) =\{f(q,p),g(q,p)\}_{PB}.
\end{equation}
This relation is the principle of correspondence. The states of the
quantized harmonic oscillator are described by the Wigner
functions $\pi_n^{(M)}(q,p)$. The Wigner functions and the energy
levels $E_n$ of the harmonic oscillator can then be calculated
with the help of the star exponential
\begin{equation}
\mathrm{Exp}_M(Ht)=e_{\starM}^{-\frac{it}{\hbar}H}=
\sum_{n=0}^{\infty}\frac{1}{n!}\left(\frac{-it}{\hbar} \right)^n
H^{n\starM}= \sum_{n=0}^{\infty}\pi_n^{(M)}e^{-iE_nt/\hbar},
\label{ExpDef}
\end{equation}
where $H^{n\starM}=H\starM\cdots\starM H$ is the $n$-fold star
product of $H$. The star exponential fulfills the analogue of the
time dependent Schr\"{o}dinger equation
\begin{equation}
i\hbar\frac{d}{dt}\mathrm{Exp}_M(Ht)=H\starM\mathrm{Exp}_M(Ht).
\end{equation}

The energy levels and the Wigner functions fulfill the $*$-genvalue equation
\begin{equation}
H\starM\pi_n^{(M)}= E_n\pi_n^{(M)}
\end{equation}
and for the harmonic oscillator one obtains
$E_n=\hbar\omega\left(n+\frac{1}{2}\right)$ and
\begin{equation}
\pi_n^{(M)}= 2(-1)^n
e^{-2H/\hbar\omega}L_n\left(\frac{4H}{\hbar\omega}\right),
\label{piMDef}
\end{equation}
where the $L_n$ are the Laguerre polynomials. 
The Wigner functions $\pi_n^{(M)}$ are normalized according to
$\frac{1}{2\pi\hbar}\int\pi_n^{(M)}\,dq\,dp=1$ and the expectation
value of a phase space function $f$ can be calculated as
\begin{equation}
\langle f\rangle=\frac{1}{2\pi\hbar}\int
f\starM\pi_n^{(M)}\,dq\,dp.
\end{equation}

The same procedure can now be used for the grassmannian case 
\cite{Deform3}. The simplest system in grassmannian mechanics
\cite{Berezin1} is a two dimensional system with Lagrange function
\begin{equation}
L=\frac{i}{2}\left(\theta_1\dot{\theta}_1+\theta_2\dot{\theta}_2\right)
+i\omega\theta_1\theta_2.
\end{equation}
With the canonical momentum
\begin{equation}
\rho_\alpha= -\frac{i}{2}\theta_{\alpha}
\label{canomo}
\end{equation}
the Hamilton function is given by
\begin{equation}
H= \dot{\theta}^{\alpha}\rho_\alpha-L= -i\omega\theta_1\theta_2.
\label{HfermDef}
\end{equation}
Together with equation (\ref{canomo}) this Hamiltonian suggests
that the fermionic oscillator describes rotation. Indeed,
calculating the fermionic angular momentum, which corresponds to
the spin, leads to
\begin{equation}
S_3=\theta_1\rho_2-\theta_2\rho_1= -i\theta_1\theta_2,
\label{Sdef}
\end{equation}
so that the Hamiltonian in (\ref{HfermDef}) can also be written as
$H=\omega S_3$. As a vector the angular momentum points out of the
$\theta_1$-$\theta_2$-plane. Therefore we consider the two
dimensional fermionic oscillator as embedded into a three
dimensional fermionic space with coordinates $\theta_1$,
$\theta_2$ and $\theta_3$. Note that we choose both for the
fermionic space and momentum coordinates the units $\sqrt{\hbar}$.

Quantizing the fermionic oscillator \cite{Deform3} involves a star
product that is given by
\begin{equation}
F\starP G= F\exp\left[\frac{\hbar}{2}\sum_{n=1}^d
\lvec{\partial}_{\theta_n}\vec{\partial}_{\theta_n}\right] G.
\label{starPDef}
\end{equation}
We will call this star product the Clifford star product because
it leads to a cliffordization of the Grassmann algebra of the 
$\theta_i$. This can be seen by considering the star-anticommutator 
that is given by
\begin{equation}
\{\theta_i,\theta_j\}_{\starP} =
\theta_i\starP\theta_j+\theta_j\starP\theta_i= \hbar\delta_{ij}.
\end{equation}
Since the Grassmann variables
\begin{equation}
\sigma^i= \frac{1}{i\hbar}\varepsilon^{ijk}\theta_j\theta_k
\textrm{\quad with\quad}i\in\{1,2,3\}, \label{sigma}
\end{equation}
fulfill the relations
\begin{equation}
\left[\sigma^i,\sigma^j\right]_{\starP}=
2i\varepsilon^{ijk}\sigma^k \qquad\mathrm{and}\qquad
\left\{\sigma^i,\sigma^j\right\}_{\starP}= 2\delta^{ij},
\end{equation}
with $\left[\sigma^i,\sigma^j\right]_{\starP}=
\sigma^i\starP\sigma^j-\sigma^j\starP\sigma^i$, they correspond to
the Pauli matrices. From
equations (\ref{Sdef}) and (\ref{sigma}) it follows that
$S_3=\frac{\hbar}{2}\sigma^3$ and $H=\omega
S_3=\frac{\hbar\omega}{2}\sigma^3$. Note, that
$\{1,\sigma^1,\sigma^2,\sigma^3\}$ is a basis of the even
subalgebra of the Grassmann algebra and that this space is also
closed under $\starP$ multiplication.

In the space of Grassmann variables there exists an analogue of
complex conjugation, which is called the involution. As in
\cite{Berezin1} it can be defined as a mapping $F \mapsto
\overline{F}$, satisfying the conditions
\begin{equation}
\overline{\overline{F}} = F \quad \text{,}\qquad \overline{F_1
F_2} = \overline{F_2}\, \overline{F_1} \qquad \text{and}\qquad
\overline{c F} = \bar c \overline{F}\text{,} \label{involution}
\end{equation}
where $c$ is a complex number and $\bar c$ its complex conjugate.
For the generators $\theta_i$ of the Grassmann algebra we assume
$\overline{\theta_i} = \theta_i$, so that for $\sigma^i$ defined
in (\ref{sigma}) the relation $\overline{\sigma^i} = \sigma^i$
holds true. This corresponds to the fact that the $2\times 2$
Pauli matrices are hermitian.

We now define the Hodge dual for Grassmann numbers with respect to
the metric $\delta_{ij}$. The Hodge dual maps a Grassmann monomial
of grade $r$ into a monomial of grade $d-r$, where $d$ is the
number of Grassmann basis elements (which is in our case three):
\begin{equation}
\star\left(\theta_{i_1}\theta_{i_2}\cdots\theta_{i_r}\right) =
\frac{1}{(d-r)!}\varepsilon^{i_{r+1}\cdots i_d}_{i_1\cdots i_r}
\theta_{i_{r+1}}\cdots\theta_{i_d}. \label{hodge}
\end{equation}
With the help of the Hodge dual one can define a trace as
\begin{equation}
\mathrm{Tr}(F)= \frac{2}{\hbar^3}\int
d\theta_3d\theta_2d\theta_1\,\star F. \label{TrDef}
\end{equation}
The integration is given by the Berezin integral for which we have
$\int d\theta_i\,\theta_j=\hbar\delta_{ij}$, where the $\hbar$ on
the right hand side is due to the fact that the variables
$\theta_i$ have units of $\sqrt{\hbar}$. The only monomial with a
non-zero trace is $1$, so that by the linearity of the integral we
obtain the trace rules
\begin{equation}
\mathrm{Tr}(\sigma^i)= 0\qquad\mathrm{and}\qquad
\mathrm{Tr}(\sigma^i\starP\sigma^j)= 2\delta^{ij} \text{.}
\label{sigma-traces}
\end{equation}

With the fermionic star product (\ref{starPDef}) one can---as in
the bosonic case---calculate the energy levels and the
$*$-eigenfunctions of the fermionic oscillator. This can be done
with the fermionic star exponential
\begin{equation}
\mathrm{Exp}_C(Ht)=e_{\starP}^{-\frac{it}{\hbar}H}
=\sum_{n=0}^{\infty}\frac{1}{n!}\left(-\frac{it}{\hbar} \right)^n
H^{n\starP} = \pi_{1/2}^{(C)}e^{-i\omega
t/2}+\pi_{-1/2}^{(C)}e^{i\omega t/2},
\end{equation}
where the Wigner functions are given by
\begin{equation}
\pi_{\pm 1/2}^{(C)}= \frac{1}{2}\mp
\frac{i}{\hbar}\theta_1\theta_2 =
\frac{1}{2}\left(1\pm\sigma^3\right) \text{.} \label{piPDef}
\end{equation}
The $\pi_{\pm1/2}^{(C)}$ fulfill the $*$-genvalue equation
$H\starP\pi_{\pm1/2}^{(C)}=E_{\pm1/2}\pi_{\pm1/2}^{(C)}$ for the
energy levels $E_{\pm1/2}=\pm\frac{\hbar\omega}{2}$. The Wigner
functions $\pi_{\pm1/2}^{(C)}$ are complete, idempotent and
normalized with respect to the trace, i.e.\ they fulfill the
equations
\begin{equation}
\pi_{+1/2}^{(C)} +\pi_{-1/2}^{(C)} = 1 %
\qquad\text{,}\qquad %
\pi_{\alpha}^{(C)}\starP \pi_{\beta}^{(C)} =
\delta_{\alpha\beta}\pi_{\alpha}^{(C)}%
\qquad\text{and}\qquad %
\mathrm{Tr} (\pi_{\pm1/2}^{(C)}) = 1,
\end{equation}
respectively. Furthermore they correspond to spin up and spin down
states since (\ref{piPDef}) corresponds to the spin projectors and
the expectation values of the angular momentum are
\begin{eqnarray}
\langle S_1\rangle&=&\mathrm{Tr}\left(\pi_{\pm1/2}^{(M)}\starP
\frac{\hbar}{2}\sigma^1\right)=0 \qquad,\qquad \langle
S_2\rangle=\mathrm{Tr}\left(\pi_{\pm1/2}^{(M)}\starP
\frac{\hbar}{2}\sigma^2\right)=0\nonumber\\
\langle S_3\rangle&=&\mathrm{Tr}\left(\pi_{\pm1/2}^{(M)}\starP
\frac{\hbar}{2}\sigma^3\right) =\pm\frac{\hbar}{2} \quad,\qquad
\langle\vec{S}^{\, 2\starP}\rangle=\mathrm{Tr}\left(
\pi_{\pm1/2}^{(M)}\starP\frac{\hbar^2}{4}\vec{\sigma}^{2\starP}
\right)=\frac{3}{4}\hbar^2.
\end{eqnarray}
where the spin $\vec{S}=\frac{\hbar}{2}\vec{\sigma}$ was used
with components of $\vec{\sigma}$ as defined in (\ref{sigma}).

In the fermionic $\theta$-space the spin
$\vec{S}=\frac{\hbar}{2}\vec{\sigma}$ is the generator of
rotations, which are described by the star exponential
\begin{equation}
\mathrm{Exp}_C(\vec{\varphi} \cdot \vec{S})=
e_{\starP}^{-\frac{1}{2}i \vec{\varphi}\cdot\vec{\sigma}}=
\cos\frac{\varphi}{2}-i(\vec{\sigma} \cdot \vec{n})
\sin\frac{\varphi}{2}, \label{RotStarExp}
\end{equation}
where we used the definition $\vec{\varphi}=\varphi \vec{n}$
with rotation angle $\varphi$ and a rotation axis given by the 
unit vector $\vec{n}$. The vector 
$\vec{\theta}=(\theta_1,\theta_2,\theta_3)^T$ transforms
passively according to
\begin{equation}
\mathrm{Exp}_C(\vec{\varphi}\cdot\vec{S})\starP \vec{\theta}
\starP\overline{\mathrm{Exp}_C(\vec{\varphi}\cdot \vec{S})}
=e_{\starP}^{-\frac{1}{2}i\vec{\varphi}
\cdot\vec{\sigma}} \starP\vec{\theta}
\starP e_{\starP}^{\frac{1}{2}i\vec{\varphi}\cdot
\vec{\sigma}} = R(\vec{\varphi})
\vec{\theta} \label{rotation}
\end{equation}
with $R(\vec{\varphi})$ being the well-known $SO(3)$
rotation matrix. The axial vector $\vec{\sigma}$
transforms in the same way. Note that the passive transformation
(\ref{rotation}) of the $\theta_i$ amounts to an active
transformation of the components $x_i$ in the vector
$x=\sum_{i=1}^3x_i\theta_i$.

\section{Geometric Algebra and the Clifford Star Product}
\setcounter{equation}{0}
Starting point for geometric algebra \cite{Hestenes1,Hestenes2} 
is an $n$-dimensional vector space over the real numbers with vectors
$\fett{a},\fett{b},\fett{c},\ldots$. A multiplication, called
geometric product, of vectors can then be denoted by juxtaposition
of an indeterminate number of vectors so that one gets monomials
$A,B,C,\ldots$. These monomials can be added in a commutative and
associative manner: $A+B=B+A$ and $(A+B)+C=A+(B+C)$, so that they
form polynomials also denoted by capital letters. The so obtained
polynomials can be multiplied associatively, i.e. $A(BC)=(AB)C$
and they fulfill the distributive laws $(A+B)C=AC+BC$ and
$C(A+B)=CA+ CB$. Furthermore there exists a null vector
$\fett{a}\fett{0}=\fett{0}$ and the multiplication with a scalar
$\lambda\fett{a}=\fett{a}\lambda$, with $\lambda\in\mathbb{R}$.
The connection between scalars and vectors can be given if one
assumes that the product $\fett{a}\fett{b}$ is a scalar iff
$\fett{a}$ and $\fett{b}$ are collinear, so that
$\sqrt{\fett{a}^2}$ is the length of the vector $\fett{a}$. These
axioms define now the Clifford algebra $C\ell(V)$ and the elements
$A,B,C,\ldots$ of $C\ell(V)$ are called Clifford or c-numbers.

Since the geometric product of two collinear vectors is a scalar,
the symmetric part of the geometric product
$\frac{1}{2}(\fett{a}\fett{b}+\fett{b}\fett{a})
=\frac{1}{2}((\fett{a}+\fett{b})^2 -\fett{a}^2-\fett{b}^2)$ is a
scalar denoted $\fett{a}\cdot \fett{b}
=\frac{1}{2}(\fett{a}\fett{b}+\fett{b}\fett{a})$. The product
$\fett{a}\cdot\fett{b}$ is the inner or scalar product. One can
then decompose the geometric product into its symmetric and
antisymmetric part:
\begin{equation}
\fett{a}\fett{b}=\frac{1}{2}(\fett{a}\fett{b}+\fett{b}\fett{a})+
\frac{1}{2}(\fett{a}\fett{b}-\fett{b}\fett{a})=\fett{a}\cdot
\fett{b}+\fett{a}\wedge \fett{b}, \label{geoprodef}
\end{equation}
where the antisymmetric part $\fett{a}\wedge
\fett{b}=\frac{1}{2}(\fett{a}\fett{b}-\fett{b}\fett{a})$ is formed
with the outer product. For the outer product one has obviously
$\fett{a}\wedge \fett{b}=-\fett{b}\wedge \fett{a}$ and
$\fett{a}\wedge \fett{a}=0$, so that $\fett{a}\wedge \fett{b}$ can
be interpreted geometrically as an oriented area. The geometric
product is constructed in such a way that it gives information
over the relative directions of $\fett{a}$ and $\fett{b}$, i.e.\
$\fett{a}\fett{b}=\fett{b}\fett{a}=\fett{a}\cdot
\fett{b}\Rightarrow \fett{a}\wedge \fett{b}=0$ means that
$\fett{a}$ and $\fett{b}$ are collinear whereas
$\fett{a}\fett{b}=-\fett{b}\fett{a}=\fett{a}\wedge
\fett{b}\Rightarrow \fett{a}\cdot \fett{b}=0$ means
that $\fett{a}$ and $\fett{b}$ are perpendicular.

With the outer product one defines simple $r$-vectors or
$r$-blades
\begin{equation}
A_r=\fett{a}_1\wedge \fett{a}_2\wedge\ldots\wedge \fett{a}_r,
\end{equation}
which can be interpreted as $r$-dimensional volume forms. The
geometric product can then be generalized to the case of a vector
and a $r$-blade:
\begin{equation}
\fett{a}A_r=\fett{a}\cdot A_r+\fett{a}\wedge A_r,
\end{equation}
which is the sum of a $(r-1)$-blade $\fett{a}\cdot
A_r=\frac{1}{2}(\fett{a}A_r -(-1)^rA_r\fett{a})$ and a
$(r+1)$-blade $\fett{a}\wedge A_r=\frac{1}{2}(\fett{a}A_r
+(-1)^rA_r\fett{a})$. Applying this recursively one sees, that
each c-number can be written as a polynomial of $r$-blades and
using a set of basis vectors
$\fett{e}_1,\fett{e}_2,\ldots,\fett{e}_r$ a c-number reads:
\begin{equation}
A=a+a^i\fett{e}_i+\frac{1}{2!}a^{i_1i_2}\fett{e}_{i_1}\wedge
\fett{e}_{i_2} +\ldots+\frac{1}{n!}a^{i_1\ldots
i_n}\fett{e}_{i_1}\wedge \fett{e}_{i_2}\wedge\ldots
\fett{e}_{i_r}.
\end{equation}
$A$ is called multivector or $r$-vector if the highest appearing
grade is $r$. It decomposes into several blades:
\begin{equation}
A=\langle A\rangle_0+\langle A\rangle_1+\ldots=\sum_n\langle
A\rangle_n,
\end{equation}
where $\langle\;\rangle_n$ projects onto the term of grade $n$. A
multivector $A_r$ is called homogeneous if all appearing blades
have the same grade, i.e.\ $A_r=\langle A_r\rangle_r$. The
geometric product of two homogeneous multivectors $A_r$ and $B_s$
can be written as
\begin{equation}
A_rB_s=\langle A_rB_s\rangle_{r+s} +\langle
A_rB_s\rangle_{r+s-2}+\cdots+\langle A_rB_s\rangle_{|r-s|}.
\end{equation}
The inner and the outer product stand now for the terms with the
lowest and the highest grade:
\begin{equation}
A_r\cdot B_s=\langle A_rB_s\rangle_{|r-s|}\qquad\mathrm{and}\qquad
A_r\wedge B_s=\langle A_rB_s\rangle_{r+s}.
\end{equation}
One should note that the inner and outer product here in the
general case do not correspond anymore to the symmetric and the
antisymmetric part of the geometric product. For example in the
case of two bivectors one has $A_2\wedge B_2=B_2\wedge A_2$, so
that the outer product is symmetric. Actually one finds for the
symmetric and the antisymmetric parts of $A_2B_2$:
\begin{equation}
\frac{1}{2}(A_2B_2+B_2A_2)=A_2\cdot B_2+A_2\wedge B_2
\qquad\mathrm{and}\qquad \frac{1}{2}(A_2B_2-B_2A_2)=\langle
A_2B_2\rangle_2.
\end{equation}

In general the commutativity of the outer and the inner product is
given by:
\begin{equation}
A_r\wedge B_s=(-1)^{rs}B_s\wedge A_r \qquad\mathrm{and}\qquad
A_r\cdot B_s=(-1)^{r(s+1)}B_s\cdot A_r
\end{equation}
and both products are always distributive:
\begin{equation}
A\wedge(B+C)=A\wedge B+A\wedge C \qquad\mathrm{and}\qquad
A\cdot(B+C)=A\cdot B+A\cdot C.
\end{equation}
Only the outer product of $r$-vectors is in general associative,
i.e.\ $A\wedge(B\wedge C)=(A\wedge B)\wedge C$, for the inner
product one gets:
\begin{equation}
A_r\cdot(B_s\cdot C_t)=(A_r\cdot B_s)\cdot C_t
\qquad\mathrm{for}\;\;\; r+t\leq s.
\end{equation}
If one has to calculate several products of different type, the
inner and the outer product always have to be calculated first,
i.e.
\begin{equation}
A\wedge BC=(A\wedge B)C\neq A\wedge(BC) \qquad\mathrm{and}\qquad
A\cdot BC=(A\cdot B)C \neq A\cdot(BC).
\end{equation}

The formalism of geometric algebra briefly sketched so far can now
be described with Grassmann variables and the Clifford star
product, that turns the Grassmann algebra into a Clifford algebra.
In order to make the equivalence even more obvious we go over to
the dimensionless Grassmann variables
\begin{equation}
\fett{\sigma}_n=\sqrt{\frac{2}{\hbar}}\theta_n. \label{normtheta}
\end{equation}
These variables play here the role of dimensionless basis vectors
and will therefore be written in bold face, whereas the $\theta_i$
played in the discussion of the first section the role of
dynamical variables with dimension $\sqrt{\hbar}$. In the
$\fett{\sigma}_n$-variables the Clifford star product
(\ref{starPDef}) has the form
\begin{equation}
F\starP
G=F\,\exp\left[\sum_{n=1}^{d}\frac{\lvec{\partial}}{\partial
\fett{\sigma}_n} \frac{\vec{\partial}}{\partial\fett{\sigma}_n}
\right]\,G. \label{starPDef2}
\end{equation}
As a star product the Clifford star product is associative and
distributive.

In order to show how the geometric algebra described with
Grassmann variables and the Clifford star product looks like, we
first consider the two dimensional euclidian case. One has then
two Grassmann basis elements $\fett{\sigma}_1$ and
$\fett{\sigma}_2$, so that a general element of the Clifford
algebra is a supernumber
$A=a_0+a_1\fett{\sigma}_1+a_2\fett{\sigma}_2
+a_{12}\fett{\sigma}_1\fett{\sigma}_2=\langle A\rangle_0+\langle
A\rangle_1+\langle A\rangle_2$ and a vector corresponds to a
supernumber with Grassmann grade one:
$\fett{a}=a_1\fett{\sigma}_1+a_2\fett{\sigma}_2$. The Clifford
star product of two of these supernumbers is
\begin{equation}
\fett{a}\starP \fett{b}=\fett{ab}+\fett{a}\left[\sum_{n=1}^2
\frac{\lvec{\partial}}{\partial\fett{\sigma}_n}
\frac{\vec{\partial}}{\partial\fett{\sigma}_n}\right]\fett{b}
=(a_1b_2-a_2b_1)\fett{\sigma}_1\fett{\sigma}_2+a_1b_1+a_2b_2\equiv
\fett{a}\wedge \fett{b}+\fett{a}\cdot \fett{b},\label{astarP2b}
\end{equation}
where the symmetric and the antisymmetric part of the Clifford
star product is given by:
\begin{eqnarray}
&&\frac{1}{2}(\fett{a}\starP \fett{b}+\fett{b}\starP \fett{a})
=a_1b_1+a_2b_2\equiv \fett{a}\cdot \fett{b}\\
\mathrm{and}&& \frac{1}{2}(\fett{a}\starP \fett{b}-\fett{b}\starP
\fett{a})=(a_1b_2-a_2b_1) \fett{\sigma}_1\fett{\sigma}_2
=\fett{ab}\equiv \fett{a}\wedge \fett{b},
\end{eqnarray}
which are terms with Grassmann grade 0 and 2 respectively. 
Note that now a juxtaposition like $\fett{ab}$ is just as in 
the notation of superanalysis the product of supernumbers and not
the Clifford product, which we want to describe explicitly 
with the star product (\ref{starPDef2}). The
$\fett{\sigma}_i$ form an orthogonal basis under the scalar
product: $\fett{\sigma}_i\cdot\fett{\sigma}_j=\frac{1}{2}
(\fett{\sigma}_i\starP\fett{\sigma}_j+\fett{\sigma}_j
\starP\fett{\sigma}_i)=\delta_{ij}$.

The unit 2-blade $\texttt{i}=\fett{\sigma}_1\fett{\sigma}_2$ can
be interpreted as the generator of $\frac{\pi}{2}$-rotations
because by multiplying from the right one gets
\begin{equation}
\fett{\sigma}_1\starP
\texttt{i}=\fett{\sigma}_1\cdot\texttt{i}=\fett{\sigma}_2
\quad,\quad \fett{\sigma}_2\starP \texttt{i}=\fett{\sigma}_2\cdot
\texttt{i}=-\fett{\sigma}_1 \quad\mathrm{and}\quad
\fett{\sigma}_1\starP \texttt{i}\starP
\texttt{i}=-\fett{\sigma}_1,
\end{equation}
so that a vector $\fett{x}=x_1\fett{\sigma}_1+x_2\fett{\sigma}_2$
is transformed into $\fett{x}'=\fett{x}\starP
\texttt{i}=\fett{x}\cdot\texttt{i}=x_1\fett{\sigma}_2
-x_2\fett{\sigma}_1$. The relation $\texttt{i}^{2\starP}=-1$
describes then a reflection and furthermore one has with
(\ref{involution}):
$\overline{\texttt{i}}=\fett{\sigma}_2\fett{\sigma}_1=-\texttt{i}$,
so that $\texttt{i}$ corresponds to the imaginary unit. The
connection between the two dimensional vector space with vectors
$\fett{x}$ and the Gauss plane with complex numbers $z$ is
established by star multiplying $\fett{x}$ with
$\fett{\sigma}_1$:
\begin{equation}
z=\fett{\sigma}_1\starP \fett{x}=x_1+ \texttt{i}x_2. \label{zsigx}
\end{equation}
Such a bivector that results from star multiplying two vectors is
also called spinor. While the bivector $\texttt{i}$ generates a
rotation of $\frac{\pi}{2}$ when acting from the right, the spinor
$z$ generates a general combination of a rotation and dilation
when acting from the right. One can see this by writing
$z=x_1+\texttt{i}x_2=|z|e_{\starP}^{\texttt{i}\varphi}$ with
$|z|^2=z\starP\overline{z}=x_1^{\,2}+x_2^{\,2}$. Acting from the
right with $z$ causes then a dilation by $|z|$ and a rotation by
$\varphi$, one has for example: $\fett{\sigma}_1\starP
z=\fett{x}$, which is the inversion of (\ref{zsigx}). Here one can
see that the formalism of geometric algebra reproduces complex
analysis and gives it a geometric meaning.

After having described the geometric algebra of the euclidian
2-space we now turn to the euclidian 3-space with basis vectors
$\fett{\sigma}_1$, $\fett{\sigma}_2$ and $\fett{\sigma}_3$ and
with the Clifford star product (\ref{starPDef2}) for $d=3$. The
basis vectors are orthogonal:
$\fett{\sigma}_i\cdot\fett{\sigma}_j=\delta_{ij}$ and a general
c-number written as a supernumber has the form
\begin{equation}
A=a_0+a_1\fett{\sigma}_1+a_2\fett{\sigma}_2+a_3\fett{\sigma}_3
+a_{12}\fett{\sigma}_1\fett{\sigma}_2
+a_{13}\fett{\sigma}_3\fett{\sigma}_1
+a_{23}\fett{\sigma}_2\fett{\sigma}_3
+a_{123}\fett{\sigma}_1\fett{\sigma}_2 \fett{\sigma}_3.
\label{A3D}
\end{equation}
This multivector has now four different simple multivector parts.
Besides the scalar part $a_0$ there is the pseudoscalar part
corresponding to
$I_3=\fett{\sigma}_1\fett{\sigma}_2\fett{\sigma}_3$, which can be
interpreted as a right handed volume form, because a parity
operation gives
$(-\fett{\sigma}_1)(-\fett{\sigma}_2)(-\fett{\sigma}_3)=-I_3$.
Moreover $I_3$ has also the properties of an imaginary unit:
$\overline{I_3}=-I_3$ and $I_3\starP I_3=I_3\cdot I_3=-1$. While
the pseudoscalar $I_3$ is an oriented volume element the bivector
part with the basic 2-blades
\begin{equation}
\texttt{i}_1=\fett{\sigma}_2\fett{\sigma}_3
=I_3\starP\fett{\sigma}_1\quad,\quad
\texttt{i}_2=\fett{\sigma}_3\fett{\sigma}_1
=I_3\starP\fett{\sigma}_2\quad\mathrm{and}
\quad\texttt{i}_3=\fett{\sigma}_1\fett{\sigma}_2
=I_3\starP\fett{\sigma}_3 \label{i123}
\end{equation}
describes oriented area elements. Each of the $\texttt{i}_r$ plays
in the plane it defines the same role as the $\texttt{i}$ of the
two dimensional euclidian plane defined above. Star-multiplying
with the pseudoscalar $I_3$ is equivalent to taking the Hodge
dual, for example to each bivector
$B=b_1\texttt{i}_1+b_2\texttt{i}_2 +b_3\texttt{i}_3$ corresponds a
vector
$\fett{b}=b_1\fett{\sigma}_1+b_2\fett{\sigma}_2+b_3\fett{\sigma}_3$,
which can be expressed by the equation $B=I_3\starP \fett{b}$.
This duality can for example be used to write the geometric
product of two vectors $\fett{a}=a_1\fett{\sigma}_1+a_2
\fett{\sigma}_2 +a_3\fett{\sigma}_3$ and
$\fett{b}=b_1\fett{\sigma}_1+b_2
\fett{\sigma}_2+b_3\fett{\sigma}_3$ as:
\begin{equation}
\fett{a}\starP \fett{b}=\fett{a}\cdot \fett{b}
+I_3\starP(\fett{a}\times \fett{b}),
\end{equation}
where $\fett{a}\cdot \fett{b}=\sum_{k=1}^{3}a_kb_k$ and
$\fett{a}\times \fett{b}=\varepsilon^{klm}a_kb_l\fett{\sigma}_m$.
Furthermore one finds:
\begin{equation}
\fett{\sigma}_1\times\fett{\sigma}_2
=-I_3\starP\fett{\sigma}_1\starP\fett{\sigma}_2
=-I_3\starP\fett{\sigma}_1\fett{\sigma}_2=\fett{\sigma}_3
\end{equation}
and cyclic permutations. Note also that one gets with the nabla
operator
$\nabla_x=\fett{\sigma}_1\partial_{x_1}+\fett{\sigma}_2\partial_{x_2}+
\fett{\sigma}_3\partial_{x_3}$ for the gradient of a vector field
$\fett{f}=f_1(x_1,x_2,x_3)\fett{\sigma}_1+f_2(x_1,x_2,x_3)\fett{\sigma}_2+
f_3(x_1,x_2,x_3)\fett{\sigma}_3$:
\begin{equation}
\nabla_x\starP
\fett{f}=\nabla_x\cdot\fett{f}+\nabla_x\wedge\fett{f}
=\mathrm{div}\,\fett{f}+I_3\starP \mathrm{rot}\, \fett{f}.
\end{equation}

The multivector part of (\ref{A3D}) with even Grassmann grade have
the basis $1,\texttt{i}_1,\texttt{i}_2,\texttt{i}_3$ and form a
closed subalgebra under the Clifford star product, namely the
quaternion algebra. The multivector part of (\ref{A3D}) with odd
grade does not close under the Clifford star product, but
nevertheless one can reinvestigate the definition of the Pauli
functions in (\ref{sigma}). Replacing in (\ref{sigma}) the scalar
$i$ by the pseudoscalar $I_3$ one sees that the basis vectors
$\fett{\sigma}_i$ fulfill
\begin{equation}
\left[\fett{\sigma}_i,\fett{\sigma}_j\right]_{\starP}
=2\varepsilon_{ijk}I_3\starP
\fett{\sigma}_k\qquad\mathrm{and}\qquad
\{\fett{\sigma}_i,\fett{\sigma}_j\}_{\starP} =2\delta_{ij},
\end{equation}
which justifies denoting them $\fett{\sigma}_i$. With the
pseudoscalar $I_3$ the trace (\ref{TrDef}) can be written as
$\mathrm{Tr}(F)=2\int
d\fett{\sigma}_3d\fett{\sigma}_2d\fett{\sigma}_1\,\star F=2\int
d\fett{\sigma}_3d\fett{\sigma}_2d\fett{\sigma}_1\, I_3\starP F$.
So one has here achieved with the Clifford star product a
cliffordization of the three dimensional Grassmann algebra of the
$\fett{\sigma}_i$.

Just as in the two dimensional case one can also consider in
three dimensions the role of spinors and rotations. To this purpose 
one first considers a vector transformation of the form
\begin{equation}
\fett{x}\rightarrow\fett{x}'=-\fett{u}\starP\fett{x}\starP\fett{u},
\label{reflect}
\end{equation}
where $\fett{u}$ is a three dimensional unit vector:
$\fett{u}=u_1\fett{\sigma}_1+u_2\fett{\sigma}_2+u_3\fett{\sigma}_3$
with $u=|\fett{u}|=\sqrt{u_1^2+u_2^2+u_3^2}=1$.  This
transformation can be identified as a reflection if one decomposes
$\fett{x}$ into a part collinear to $\fett{u}$ and a part
orthogonal to $\fett{u}$:
\begin{equation}
\fett{x}=\fett{x}_{\|}+\fett{x}_{\bot}=(\fett{x}\cdot
\fett{u}+\fett{x}\fett{u})\starP \fett{u},
\end{equation}
with $\fett{x}_{\|}=(\fett{x}\cdot \fett{u})\fett{u}$ and
$\fett{x}_{\bot}=(\fett{x}\fett{u})\starP
\fett{u}=(\fett{x}\fett{u})\cdot \fett{u}$. One can easily check
that
\begin{equation}
\fett{x}_{\|}\starP \fett{u}=\fett{u}\starP
\fett{x}_{\|}\Rightarrow \fett{x}_{\|}\| \fett{u}
\qquad\mathrm{and}\qquad \fett{x}_{\bot}\starP
\fett{u}=-\fett{u}\starP \fett{x}_{\bot}\Rightarrow
\fett{x}_{\bot}\bot \fett{u}. \label{xparasenk}
\end{equation}
This decomposition of $\fett{x}$ can most easily be obtained if
one just star-divides $\fett{x}\starP \fett{u}=\fett{x}\cdot
\fett{u}+\fett{x}\wedge \fett{u}$ by $\fett{u}$, which gives with
$\fett{u}^{-1\starP}=\fett{u}$:
\begin{equation}
\fett{x}=(\fett{x}\cdot \fett{u})\starP
\fett{u}^{-1\starP}+(\fett{x}\fett{u})\starP
\fett{u}^{-1\starP}=(\fett{x}\cdot
\fett{u})\fett{u}+(\fett{x}\fett{u})\starP
\fett{u}=\fett{x}_{\|}+\fett{x}_{\bot}.\label{uinvers}
\end{equation}
Using (\ref{xparasenk}) one sees that the transformation
(\ref{reflect}) turns $\fett{x}$ into
$\fett{x}'=-\fett{u}\starP\fett{x}\starP\fett{u}
=-\fett{x}_{\|}+\fett{x}_{\bot}$, so that only the component
collinear to $\fett{u}$ is inverted, which amounts to a reflection
at the plane where $\fett{u}$ is the normal vector. Two successive
transformations (\ref{reflect}) lead to:
\begin{equation}
\fett{x}\rightarrow \fett{x}''=-\fett{v}\starP \fett{x}'\starP
\fett{v}=\fett{v}\starP \fett{u}\starP \fett{x}\starP
\fett{u}\starP \fett{v}=U\starP \fett{x}\starP\overline{U},
\label{UxU}
\end{equation}
where $U$ can be written as:
\begin{equation}
U=\fett{v}\starP \fett{u}=\fett{v}\cdot \fett{u}+\fett{v}\wedge
\fett{u}=\cos\left(\frac{1}{2}|A|\right)+A_0\sin\left(\frac{1}{2}
|A|\right)=e_{\starP}^{\frac{1}{2}A}, \label{Udo}
\end{equation}
where the angle between the unit vectors $\fett{u}$ and $\fett{v}$
is described by an bivector $A=\fett{v}\wedge
\fett{u}=\fett{v}\fett{u}=|\fett{v}\fett{u}|A_0$. Hereby the unit
bivector $A_0=\fett{v}\fett{u}/|\fett{v}\fett{u}|$ defines the
plane in which the angle lies, while the magnitude
$|\fett{v}\fett{u}|$ gives the angle in radians, furthermore it fulfills
$A_0\starP A_0=-1$. If one chooses for example the basis vectors
$\fett{\sigma}_k$ for $\fett{u}$ and $\fett{v}$, $A_0$ would be
given by one of the bivectors in (\ref{i123}). The additional
factor $1/2$ in (\ref{Udo}) will become clear if one investigates
the action of the transformation (\ref{UxU}). To this purpose one
proceeds analogous to the discussion of the reflection
(\ref{reflect}). One first decomposes the vector $\fett{x}$ into a
part $\fett{x}_{\|}$ in the plane defined by $A$ and a part
$\fett{x}_{\bot}$ perpendicular to that plane. This is done
analogous to (\ref{uinvers}) by star-dividing $\fett{x}\starP
A=\fett{x}\cdot A+\fett{x}\wedge A$ by $A$ which leads to
\begin{equation}
\fett{x}=(\fett{x}\cdot A)\starP A^{-1\starP}+(\fett{x}A)\starP
A^{-1\starP}=\fett{x}_{\|}+\fett{x}_{\bot},
\end{equation}
with $\fett{x}_{\|}\starP A=-A\starP \fett{x}_{\|}$ and
$\fett{x}_{\bot}\starP A=A\starP \fett{x}_{\bot}$. We then have
for the transformation (\ref{UxU}):
\begin{equation}
U\starP \fett{x}\starP\overline{U}=e_{\starP}^{-A/2}\starP
\fett{x}\starP e_{\starP}^{A/2}
=\fett{x}_{\bot}+\fett{x}_{\|}\starP e_{\starP}^A.
\label{xrotation}
\end{equation}
So the component perpendicular to the plane defined by $A$ is not
changed while the component inside this plane is rotated in that
plane with the help of the spinor $e_{\starP}^A$ by an angle of
magnitude $|A|$, just as described in the two dimensional
case above. One sees here why the rotation in the two dimensional
case could be written just by acting with a spinor from the right.
This is due to the fact that when the vector lies in the plane of
rotation one has
\begin{equation}
e_{\starP}^{-A/2}\starP \fett{x}_{\|}\starP
e_{\starP}^{A/2}=\fett{x}_{\|} \starP e_{\starP}^A.
\end{equation}
A rotation can be described with the bivector $A$, but also with
the dual vector $\fett{a}$ defined by $A=I_3\starP\fett{a}$, where
the direction of $\fett{a}$ defines the axis of rotation, while
the magnitude gives the radian $|\fett{a}|=|A|$. So $U$ can also
be written as:
\begin{equation}
U=e_{\starP}^{-\frac{1}{2}I_3\starP{\mbox{\footnotesize\boldmath$a$}}},
\end{equation}
which corresponds to the star exponential (\ref{RotStarExp}).

The formalism described so far can easily be generalized to the
case of $d$ euclidian dimensions. Just as there is a duality
inside the space spanned by the $\fett{\sigma}_i$ there is also
the duality between the spaces spanned by the $\fett{\sigma}_i$
and the $\fett{\sigma}^i$. This duality is expressed by the
relation $\fett{\sigma}_i\cdot \fett{\sigma}^j=\delta^j_i$. The
$\fett{\sigma}^i$-vectors can be constructed with the help of the
pseudoscalar, which is for the $d$-dimensional euclidian case
$I_d=\fett{\sigma}_1 \fett{\sigma}_2\ldots\fett{\sigma}_d$. The
space on which the basis vector $\fett{\sigma}_j$ is normal is
given for an $d$-dimensional euclidian space by the $(d-1)$-blade
$(-1)^{j-1}
\fett{\sigma}_1\fett{\sigma}_2\ldots\check{\fett{\sigma}}_j\ldots
\fett{\sigma}_d$, where $\check{\fett{\sigma}}_j$ means that this
basis vector is missing. The corresponding dual vector is then
given by
\begin{equation}
\fett{\sigma}^j=(-1)^{j-1}\fett{\sigma}_1\fett{\sigma}_2
\ldots\check{\fett{\sigma}}_j\ldots \fett{\sigma}_d\starP
I_d^{-1\starP},\label{obendual}
\end{equation}
where $I_d^{-1\starP}$ is the inverse $d$-dimensional
pseudoscalar.

Note also that the multiple Clifford star product leads to an 
expansion of Wick type. For example the Clifford product of
four basis vectors is given by
\begin{eqnarray}
\fett{\sigma}_{i_1}\starP\fett{\sigma}_{i_2}
\starP\fett{\sigma}_{i_3}\starP\fett{\sigma}_{i_4}&=&
\fett{\sigma}_{i_1}\fett{\sigma}_{i_2}\fett{\sigma}_{i_3}\fett{\sigma}_{i_4}
+\fett{\sigma}_{i_1}\fett{\sigma}_{i_2}\delta_{i_3i_4}
-\fett{\sigma}_{i_1}\fett{\sigma}_{i_3}\delta_{i_2i_4}
+\fett{\sigma}_{i_1}\fett{\sigma}_{i_4}\delta_{i_2i_3}\nonumber\\
&&+\fett{\sigma}_{i_2}\fett{\sigma}_{i_3}\delta_{i_1i_4}
-\fett{\sigma}_{i_2}\fett{\sigma}_{i_4}\delta_{i_1i_3}
+\fett{\sigma}_{i_3}\fett{\sigma}_{i_4}\delta_{i_1i_2}\nonumber\\
&&+\delta_{i_1i_2}\delta_{i_3i_4}
-\delta_{i_1i_3}\delta_{i_2i_4}
+\delta_{i_1i_4}\delta_{i_2i_3},
\end{eqnarray}
where the contraction of $\fett{\sigma}_i$ and $\fett{\sigma}_j$ is given
by $\delta_{ij}$. This suggests to use the star product formalism also in 
the realm of quantum field theory \cite{pert}.

\section{Geometric Algebra and Classical Mechanics}
\setcounter{equation}{0}
It is now straight forward to use the formalism described so far
in classical mechanics as it was done in \cite{Hestenes2}. We will
here only give two examples to show where the advantages of
geometric algebra lie. Let us first consider the three dimensional
harmonic oscillator, which is defined by the differential equation
$\ddot{\fett{q}}+\frac{k}{m}\fett{q}=0$, where $\fett{q}$ is now a
supernumber: $\fett{q}=q_1\fett{\sigma}_1
+q_2\fett{\sigma}_2+q_3\fett{\sigma}_3$. The ansatz
$\fett{q}=\fett{a}\starP e_{\starP}^{\lambda t}$ leads to the
equation $\lambda^{2\starP}+\frac{k}{m}=0$, which is solved by
$\lambda=\pm \texttt{i}\omega_0$ with $\omega_0=\sqrt{k/m}$. The
difference to the conventional formalism is that $\texttt{i}$ is
here a bivector with $\texttt{i}^{2\starP}=-1$. This gives then
the two solutions
\begin{equation}
\fett{q}_{\pm}=\fett{a}_{\pm}\starP e_{\starP}^{\pm\texttt{i}
\omega_0t} =\fett{a}_{\pm}\starP (\cos\omega_0t\pm
\texttt{i}\sin\omega_0t).
\end{equation}
In the second term appears the expression
$\fett{a}_{\pm}\starP\texttt{i}
=\fett{a}_{\pm}\texttt{i}+\fett{a}_{\pm}\cdot \texttt{i}$, which
is the sum of a term of Grassmann grade three and a term of
Grassmann grade one. But the result $\fett{q}_{\pm}$ itself is a
quantity of Grassmann grade one, so it follows that
$\fett{a}_{\pm}\texttt{i}=0$, which is the defining equation of
the plane in which the oscillatory movement takes place. This
plane is defined by the unit bivector $\texttt{i}$ and has to be
determined by the initial conditions \cite{Hestenes2}.

As the second example we consider the solution of the Kepler
problem by spinors \cite{Hestenes8}. One uses here the fact that
the radial position vector $\fett{r}=r_1\fett{\sigma}_1+
r_2\fett{\sigma}_2 +r_3\fett{\sigma}_3$ can be written as a
rotated and dilated basis vector:
\begin{equation}
\fett{r}=U\starP\fett{\sigma}_1\starP\overline{U}.\label{rUsigU}
\end{equation}
The components $r_i$ of $\fett{r}$ can then be expressed in terms
of the components $u_i$ of
$U=u_1+u_2\fett{\sigma}_2\fett{\sigma}_3 +u_3 \fett{\sigma}_3
\fett{\sigma}_1+u_4\fett{\sigma}_1 \fett{\sigma}_2$:
\begin{equation}
\left(\begin{array}{c}
  r_1 \\
  r_2 \\
  r_3 \\
  0 \\
\end{array}\right)
=\left(\begin{array}{rrrr}
  u_1 & u_2 & -u_3 & -u_4 \\
  -u_4 & u_3 & u_2 & -u_1 \\
  u_3 & u_4 & u_1 & u_2 \\
  -u_2 & u_1 & -u_4 & u_3 \\
\end{array}\right)
\left(\begin{array}{c}
  u_1 \\
  u_2 \\
  u_3 \\
  u_4 \\
\end{array}\right),\label{rLu}
\end{equation}
which is the well known Kustaanheimo-Stiefel transformation
\cite{KS,Stiefel}. Comparing (\ref{rUsigU}) and (\ref{rLu}) leads to the
notational correspondence
\begin{equation}
\fett{r}=U\starP\fett{\sigma}_1\starP\overline{U}
\qquad\leftrightarrow\qquad \vec{r}=L_{\vec{u}}\,\vec{u},
\label{KStrafbeid}
\end{equation}
where $\vec{r}$ and $\vec{u}$ are four dimensional space vectors
considered as tupels of numbers as in the conventional formalism.
One should note here that the KS-transformation increases the
degrees of freedom by one, which means that the bivector $U$ in
(\ref{rUsigU}) is not unique \cite{Hestenes8}. This gauge freedom
can be reduced by imposing an additional constraint on $U$ as will
be shown below. Squaring (\ref{KStrafbeid}) leads to the relations
\begin{equation}
\;\;\;\;\;\;\;\;\;\;
U\starP\overline{U}=|U|^2=r\qquad\leftrightarrow\qquad
L_{\vec{u}}L^T_{\vec{u}}=\vec{u}^{\,2}=r,
\end{equation}
with $r=|\fett{r}|=|\vec{r}\,|=r_1^2+r_2^2+r_3^2$. Differentiating
(\ref{KStrafbeid}) with respect to $t$ one obtains the
KS-transformation for the velocities as
\begin{equation}
\!\!\!\!\!\!\!\!\!\!\!\!\!\!\!\!\!\!\!\!\!\!\!\!\!\!\!\!\!\!\!
\!\!\!\!\!\!\!\!
\dot{\fett{r}}=\dot{U}\starP\fett{\sigma}_1\starP\overline{U}
+U\starP\fett{\sigma}_1\starP\dot{\overline{U}}
\qquad\leftrightarrow\qquad
\dot{\vec{r}}=2L_{\vec{u}}\,\dot{\vec{u}}. \label{rdotUdottraf}
\end{equation}
One can then choose for the constraint
\begin{equation}
\!\!\!\!\!\!\!\!\!\!\!\!\!\!\!\!\!\!\!\!\!\!\!\!\!\!\!\!\!\!\!
\!\!\!
\dot{U}\starP\fett{\sigma}_1\starP\overline{U}=U\starP
\fett{\sigma}_1\starP
\dot{\overline{U}} \qquad\leftrightarrow\qquad \dot{r}_4=0,
\label{Uconstraint}
\end{equation}
which means that the superfluous fourth component $r_4$ stays zero for
all times. With this constraint it is possible to invert the
geometric algebra relation (\ref{rdotUdottraf}) for $U$.
Implementing (\ref{Uconstraint}) in (\ref{rdotUdottraf}) gives
$\dot{\fett{r}}=2\dot{U}\starP\fett{\sigma}_1 \starP\overline{U}$,
which can be solved for $\dot{U}$, so that the inverse relation to
(\ref{rdotUdottraf}) is
\begin{equation}
\dot{U}=\frac{1}{2r}\dot{\fett{r}}\starP U\starP\fett{\sigma}_1
\qquad\leftrightarrow\qquad
\dot{\vec{u}}=\frac{1}{2r}L_{\vec{u}}^T\dot{\vec{r}}.
\label{Upunktr}
\end{equation}

By introducing a fictitious time $s$ which is defined as
\begin{equation}
\frac{d}{ds}=r\frac{d}{dt},\qquad \frac{dt}{ds}=r \label{tfikti}
\end{equation}
it is then possible to regularize the divergent $1/r$-potential so
that (\ref{Upunktr}) reads
$\frac{dU}{ds}=\frac{1}{2}\dot{\fett{r}}\starP
U\starP\fett{\sigma}_1$ or
\begin{equation}
\frac{d^2U}{ds^2}=\frac{1}{2}\left(r\,\ddot{\fett{r}}\starP
U\starP\fett{\sigma}_1
+\dot{\fett{r}}\starP\frac{dU}{ds}\starP\fett{\sigma}_1\right)=
\frac{1}{2}U\starP\left(\ddot{\fett{r}}\starP\fett{r}+\frac{1}{2}
\dot{\fett{r}}^{2\starP} \right).
\end{equation}
Substituting now the inverse square force
\begin{equation}
m\ddot{\fett{r}}=-k\frac{\fett{r}}{r^3}\label{inverssquareforce}
\end{equation}
one obtains:
\begin{equation}
\frac{d^2U}{ds^2}=\frac{1}{2m}U\starP\left(\frac{1}{2}m
\dot{\fett{r}}^{2\starP} -\frac{k}{r}\right)=\frac{E}{2m}U,
\end{equation}
which is the equation of motion for an harmonic oscillator. This
equation can be solved in a straight forward fashion and is much easier 
than the equation for $\fett{r}$. The orbit can then be calculated by
(\ref{rUsigU}).

The Kepler problem can also be treated in the canonical formalism.
Therefore one first needs the KS-transformation for the momentum.
If $\fett{w}=\sum_{n=1}^{4} w_n\fett{\sigma}_n$ is the canonical
momentum corresponding to
$\fett{u}=\sum_{n=1}^{4}u_n\fett{\sigma}_n$ the KS-transformation
is given by
\begin{equation}
\fett{p}=\frac{1}{4r}\left(W\starP\fett{\sigma}_1
\starP\overline{U}+U\starP\fett{\sigma}_1
\starP\overline{W}\right) \qquad\leftrightarrow\qquad
\vec{p}=\frac{1}{2r}L_{\vec{u}}\vec{w}, \label{pwtrafo}
\end{equation}
with $W=w_1+w_2\fett{\sigma}_2\fett{\sigma}_3
+w_3\fett{\sigma}_3\fett{\sigma}_1
+w_4\fett{\sigma}_1\fett{\sigma}_2$. For
$\fett{p}^{2\starP}=p_1^2+p_2^2+p_3^2$ one gets with
(\ref{pwtrafo})
\begin{equation}
\fett{p}^{2\starP}=\frac{1}{4r}|W|^2-p_4^2, \label{pquadrat}
\end{equation}
where $|W|^2=W\starP\overline{W}=w_1^2+w_2^2+w_3^2+w_4^2$ and
\begin{equation}
p_4=\frac{1}{2r}\left(u_1w_2-u_2w_1+u_3w_4-u_4w_3\right).
\label{pvierDef}
\end{equation}
Equation (\ref{pquadrat}) allows to transform the Hamiltonian into
$u_i$- and $w_i$-coordinates. This is done in several steps
\cite{Stiefel}. Starting from the Hamiltonian
$H=\frac{1}{2m}(p_1^2+p_2^2+p_3^2)-\frac{k}{r}$ one first extends
the phase space by a $q_0$- and a $p_0$-coordinate and forms the
homogenous Hamiltonian as $H_1=H+p_0$. This leads for the zero
component to two additional Hamilton equations
\begin{equation}
\frac{dq_0}{dt}=\frac{\partial H_1}{\partial p_0}=1
\qquad\mathrm{and}\qquad \frac{dp_0}{dt}=-\frac{\partial
H_1}{\partial q_0}=-\frac{\partial H_1}{\partial t}
=-\frac{\partial p_0}{\partial t},
\end{equation}
which shows that $q_0$ corresponds to the time $t$ and $p_0$ is a
constant and corresponds to the negative energy of the system, so
that $H_1=H+p_0=0$ for a conservative force. Since the time is now
a coordinate the development of the system has to be described
with a different parameter. This development parameter is the
fictitious time $s$ that is connected to the time by
(\ref{tfikti}). The relation (\ref{tfikti}) can be implemented if
one chooses $H_2=rH_1$. The Hamilton equations that describe then
the development according to $s$ are differential equations with
respect to $s$:
\begin{equation}
\frac{dq_i}{ds}=\frac{\partial H_2}{\partial p_i}
\qquad\mathrm{and}\qquad \frac{dp_i}{ds}=-\frac{\partial
H_2}{\partial q_i}\qquad\mathrm{for}\;\;i=0,1,2,3.
\end{equation}
Especially for the zero component one gets $\frac{dq_0}{ds}
=\frac{dt}{ds}=\frac{\partial H_2}{\partial p_0}=r$ which
corresponds to (\ref{tfikti}). After having so far regularized the
Hamiltonian one can then go over to KS-coordinates and obtains
with (\ref{pquadrat})
\begin{equation}
H_3=\frac{1}{8m}\left(w_1^2+w_2^2+w_3^2+w_4^2\right)
-\frac{1}{2m}rp_4^2-k-Er.
\end{equation}
Imposing now the constraint $p_4=0$, which for $w_i=m\dot{u}_i$ in
(\ref{pvierDef}) is just (\ref{Uconstraint}), and considering bound 
states with $E<0$ the Hamiltonian is given by
\begin{equation}
H_4=\frac{1}{8m}\left(w_1^2+w_2^2+w_3^2+w_4^2\right)
+|E|\left(u_1^2+u_2^2+u_3^2+u_4^2\right)-k,\label{Hvier}
\end{equation}
which describes a four dimensional harmonic oscillator with fixed
energy and frequency $\omega=(|E|/2m)^{1/2}$.

The formalism of geometric algebra can also be applied to
hamiltonian mechanics \cite{Hestenes9}. The $2d$-dimensional phase
space is then spanned by $d$ basis elements $\{\fett{\eta}_i\}$
for the space coordinates and $d$ basis elements
$\{\fett{\rho}_i\}$ for the momentum coordinates so that a point
in phase space is described by the vector $\fett{x}=\sum_{n=1}^d
(q_n\fett{\eta}_n+p_n\fett{\rho}_i)$ and the Clifford star product
on the phase space has the from
\begin{equation}
F\starP G =F\,\exp\left[\sum_{n=1}^{d}\left(\frac{\lvec{\partial}}
{\partial\fett{\eta}_n}\frac{\vec{\partial}}{\partial\fett{\eta}_n}+
\frac{\lvec{\partial}}{\partial\fett{\rho}_n}
\frac{\vec{\partial}}{\partial\fett{\rho}_n}\right)\right]\,G,
\label{starPDef3}
\end{equation}
so that $\fett{\eta}_m\cdot\fett{\eta}_n =\fett{\rho}_m
\cdot\fett{\rho}_n=\delta_{mn}$ and
$\fett{\eta}_m\cdot\fett{\rho}_n=0$. The two $d$-dimensional
subspaces are related by a bivector $\texttt{j}$, which is the
generalization of the imaginary structure in two dimensions and is
defined as:
\begin{equation}
\texttt{j} =\sum_{n=1}^d \texttt{j}_n=
\sum_{n=1}^{d}\fett{\eta}_n\fett{\rho}_n. \label{Jsigsig}
\end{equation}
This bivector plays the role of the symplectic form that relates
the space and momentum part of the phase space according to
\begin{equation}
\fett{\eta}_n\cdot\texttt{j}=\fett{\rho}_n
\qquad\mathrm{and}\qquad
\texttt{j}\cdot\fett{\rho}_n=\fett{\eta}_n.
\end{equation}
In phase space one has then two possibilities to assign a scalar
to two phase space vectors $\fett{a}$ and $\fett{b}$, apart from
the scalar product $\fett{a}\cdot\fett{b}$ one can also form the
expression
\begin{equation}
\fett{a}\cdot(\texttt{j}\cdot\fett{b})=-\texttt{j}\cdot
(\fett{a}\fett{b})\equiv \fett{a}\cdot\widetilde{\fett{b}}.
\end{equation}
With the gradient operator
\begin{equation}
\nabla_x=\sum_{n=1}^d\left(\fett{\eta}_n\frac{\partial}{\partial
q_n} +\fett{\rho}_n\frac{\partial}{\partial p_n}\right)
\label{nablax}
\end{equation}
the Hamilton equation can for example be written as:
\begin{equation}
\dot{\fett{x}}=\texttt{j}\cdot(\nabla_xH)=\widetilde{\nabla}_xH,
\label{HamilJ}
\end{equation}
or explicitly:
\begin{equation}
\sum_{n=1}^d\left(\dot{q}_n\fett{\eta}_n
+\dot{p}_n\fett{\rho}_n\right)
=\texttt{j}\cdot\sum_{n=1}^d\left(\fett{\eta}_n\frac{\partial
H}{\partial q_n} +\fett{\rho}_n\frac{\partial H}{\partial
p_n}\right) =\sum_{n=1}^d\left(-\fett{\rho}_n\frac{\partial
H}{\partial q_n}+\fett{\eta}_n\frac{\partial H}{\partial
p_n}\right).
\end{equation}
With (\ref{HamilJ}) one gets for the time derivation of a
scalar phase space function $f(x)$:
\begin{equation}
\dot{f}=\dot{\fett{x}}\cdot(\nabla_xf)
=\sum_{n=1}^d\left(\frac{\partial f}{\partial q_n}\frac{\partial
H}{\partial p_n}-\frac{\partial H}{\partial q_n}\frac{\partial
f}{\partial p_n}\right)=\{f,H\}_{PB}.
\end{equation}
The Poisson bracket can be written in a compact way as:
\begin{equation}
\{f,g\}_{PB}=f\left(-\texttt{j}\cdot\left(\lvec{\nabla}_x
\vec{\nabla}_x\right)\right)g=f\left(\lvec{\nabla}_x\cdot
\widetilde{\vec{\nabla}}_x\right)g. \label{PBnabnab}
\end{equation}

\section{Nonrelativistic Quantum Mechanics}
\setcounter{equation}{0}
The above discussed transformation of the Kepler problem can now
be used to calculate the energy levels of the hydrogen atom as it 
was described in \cite{Gracia-Bondia}.
To this purpose one introduces holomorphic coordinates
\begin{equation}
a_n=\frac{1}{\sqrt{2}}\left(\sqrt{4m\omega}\,u_n+i
\frac{1}{\sqrt{4m\omega}}w_n\right)
\end{equation}
so that the Hamiltonian $H_4$ in (\ref{Hvier}) can be written as:
\begin{equation}
H_4=\omega\left(\sum_{n=1}^4a_n \bar{a}_n\right)-e^2, \label{H42}
\end{equation}
where $k=e^2$. Introducing then holomorphic coordinates for left
and right moving quanta 
\begin{equation}
a_{R_{12}}=\frac{1}{\sqrt{2}}\left(a_1-i
a_2\right),\quad a_{L_{12}}=\frac{1}{\sqrt{2}}\left(a_1+i a_2\right)
\quad\mathrm{and}\quad a_{R_{34}}=\frac{1}{\sqrt{2}}\left(a_3-i a_4\right),
\quad a_{L_{34}}=\frac{1}{\sqrt{2}}\left(a_3+i a_4\right)
\end{equation}
the Hamiltonian (\ref{H42}) turns into
\begin{equation}
H_4=\omega\left(a_{R_{12}}\bar{a}_{R_{12}}
+a_{L_{12}}\bar{a}_{L_{12}}+ a_{R_{34}}\bar{a}_{R_{34}}
+a_{L_{34}}\bar{a}_{L_{34}}\right)-e^2.
\end{equation}
One can now quantize this system with the Moyal product.  The four
dimensional Moyal star product transforms under KS-transformation 
and the above transformations into
\begin{equation}
\starM = \exp\Bigg[\sum_{n=1}^{4}\frac{i\hbar}{2}
\left(\lvec{\partial}_{u_n}\vec{\partial}_{w_n}-
\lvec{\partial}_{w_n}\vec{\partial}_{u_n}\right)\Bigg] = \exp\left[
\frac{\hbar}{2}\sum_{X=R_{12},L_{12},R_{34},L_{34}}
\left(\lvec{\partial}_{a_X}\vec{\partial}_{\bar{a}_X}
-\lvec{\partial}_{\bar{a}_X}\vec{\partial}_{a_X}\right)\right].
\end{equation}
The energy levels can then be obtained by the $*$-genvalue
equation
\begin{equation}
H_4\starM \pi_{n_1n_2n_3n_4}^{(M)}=0,\label{H4stargen}
\end{equation}
where $\pi_{n_1n_2n_3n_4}^{(M)}$ is the product of four Wigner
functions of the one dimensional harmonic oscillator given in
(\ref{piMDef}). Eq.\ (\ref{H4stargen}) gives then
\begin{equation}
e^2=\hbar\omega\left(n_{R_{12}}+n_{L_{12}}+n_{R_{34}}+n_{L_{34}}
+2\right). \label{ekm}
\end{equation}
To get the energy levels of the hydrogen atom one has to impose
the constraint
\begin{equation}
p_4=a_{R_{12}}\bar{a}_{R_{12}}
-a_{L_{12}}\bar{a}_{L_{12}}+a_{R_{34}}\bar{a}_{R_{34}}
-a_{L_{34}}\bar{a}_{L_{34}}=0,
\end{equation}
which for the energy levels corresponds to
$n_{R_{12}}-n_{L_{12}}+n_{R_{34}}-n_{L_{34}}=0$ or
$n_{R_{12}}+n_{R_{34}}$$=n_{L_{12}}+n_{L_{34}}$$\equiv n-1$. Putting
this and $\omega=\sqrt{|E|/2m}$ into (\ref{ekm}) one gets the well
known energy levels of the hydrogen atom
\begin{equation}
E_n=-\frac{e^4m}{2\hbar}\frac{1}{n^2}.
\end{equation}

Geometric algebra in a fermionic star product formalism is not
only useful for calculating the energy levels of the hydrogen
atom, it can also be combined straight forwardly with the bosonic
star product formalism of quantum mechanics. In classical
mechanics described with geometric algebra and the Clifford star
product the fermionic part of the underlying superanalysis was
deformed and the basis vectors played only a mathematical role by
generating the structures of vector analysis. Going over to
quantum mechanics means that also the scalar coefficients of
superanalysis have to be multiplied by a deformed product, namely
the bosonic Moyal star product. This leads then to a deformed
version of geometric algebra and describing geometric algebra in terms
of star products allows to combine the Clifford star product
and the Moyal product into one star product, which should be called 
Moyal-Clifford product. The Clifford product on the phase
space that described the structures of classical Hamilton mechanics 
was given by (\ref{starPDef3}). In quantum mechanics one needs now a 
product with which general multivector functions on the phase space are 
multiplied. These multivector functions are the observables of the 
theory and as such can only be multivectors in the space basis vectors
$\fett{\sigma}_r$. So one has to go over from the Clifford product
(\ref{starPDef3}) to the Clifford product (\ref{starPDef2}), which
can be done by implementing constraints that identify the corresponding
basis vectors \cite{Deform3}. The Moyal-Clifford product for a 
single particle system is then    
\begin{equation}
F\starMP G
=F\,\exp\left[\sum_{n=1}^{d}\left(\frac{i\hbar}{2}(\lvec{\partial}_{q_n}
\vec{\partial}_{p_n}-\lvec{\partial}_{p_n}\vec{\partial}_{q_n})
+\lvec{\partial}_{{\mbox{\footnotesize\boldmath$\sigma$}}_n}
\vec{\partial}_{{\mbox{\footnotesize\boldmath$\sigma$}}_n}\right)\right]
\,G.
\end{equation}
To see the consequences of the additional Moyal deformation in geometric
algebra one can for example consider the Moyal-Clifford product of two 
vectors in $d=2$ dimensions. The generalization of (\ref{astarP2b}) 
can be written as
\begin{equation}
\fett{a}\starMP \fett{b}
=(a_1\starM b_2-a_2\starM b_1)\fett{\sigma}_1\fett{\sigma}_2
+a_1\starM b_1+a_2\starM b_2.
\end{equation}
Under the Moyal product the coefficients in general do not commute if 
they are functions of $q_n$ and $p_n$. This means that the 
Moyal-Clifford product of the same vectors $\fett{a}\starMP\fett{a}$
is in general not a scalar, but has also a bivector part. It is 
this additional bivector part, which appears only for $\hbar\neq 0$, that
constitutes the spin as a physical observable. This can be seen if
one considers the minimal substituted Hamiltonian which
is in the formalism of deformed geometric algebra given by:
\begin{eqnarray}
H&=&\frac{1}{2m}\Big[\left(p_1+eA_1\right)\fett{\sigma}_1+
\left(p_2+eA_2\right)\fett{\sigma}_2+
\left(p_3+eA_3\right)\fett{\sigma}_3\Big]^{2\starMP}\label{Hqua1}\\
&=&\frac{1}{2m}\Big[\left(p_1+eA_1\right)^{2\starM}+
\left(p_2+eA_2\right)^{2\starM}+
\left(p_3+eA_3\right)^{2\starM}\Big]\nonumber\\
&&+\frac{1}{2m}\Big[\left(p_1+eA_1\right),
\left(p_2+eA_2\right)
\Big]_{\starM}\fett{\sigma}_1\fett{\sigma}_2
+\frac{1}{2m}\Big[\left(p_1+eA_1\right),
\left(p_3+eA_3\right)\Big]_{\starM}
\fett{\sigma}_1\fett{\sigma}_3\nonumber\\
&&+\frac{1}{2m}\Big[\left(p_2+eA_2\right),
\left(p_3+eA_3\right) \Big]_{\starM}
\fett{\sigma}_2\fett{\sigma}_3.\label{Hqua2}
\end{eqnarray}
The first three terms $H_0=\frac{1}{2m}\sum_{n=0}^3\left(p_n
+eA_n\right)^{2\starM}$ describe the Landau problem of a
charged particle in a magnetic field which can be solved in the
star product formalism as described in \cite{Demircioglu} or
\cite{Deform5}. The other three terms that describe the
interaction of the spin and the magnetic field appear only because
of the Moyal product. If the magnetic field points in
$\fett{\sigma}_3$-direction the vector potential is given by
$\fett{A}=-\frac{B_3}{2}q_2\fett{\sigma}_1
+\frac{B_3}{2}q_1\fett{\sigma}_2$ and only the first
Moyal-commutator contributes:
\begin{equation}
H_S=\frac{1}{2m}\Big[\left(p_1+eA_1\right),
\left(p_2+eA_2\right)
\Big]_{\starM}\fett{\sigma}_1\fett{\sigma}_2
=\frac{\hbar\omega}{2}\sigma^3\label{HSDef}
\end{equation}
where $\omega=\frac{eB_3}{m}$ and
$\sigma^3=-i\fett{\sigma}_1\fett{\sigma}_2$ is a real quaternion,
which is constructed according to (\ref{normtheta}) and
(\ref{sigma}). The difference between this calculation and the
conventional approach is that in the conventional formalism the
Clifford structure is introduced by putting in Pauli
matrices by hand in (\ref{Hqua1}). The Pauli matrices describe
the spin and lead analogously to the additional term
$H_S$, this approach is also known as the Feynman trick 
\cite{Sakurai}. In
geometric algebra the Clifford structures do not have to be added,
they are just the basis vectors that already exist in classical
mechanics, but become apparent as physical objects in the quantum
case. It is then straight forward to calculate the
$*$-eigenfunctions of $H_S$ which turn out to be the spin Wigner
functions described in the first section \cite{Deform5}.

One should note that the Moyal-Clifford product is a product for
functions on the phase space, which play the role of observables.
As seen above these observables are in general multivectors, where
the terms of higher grade are described by the space basis vectors 
$\fett{\sigma}_n$ and not by the phase space basis vectors 
$\fett{\eta}_n$ and $\fett{\rho}_n$, because the latter should not be
observable quantities. Nevertheless the basis vectors of phase 
space can be considered to play an indirect role in the   
expression (\ref{starMDef}) of the Moyal product, because the imaginary
structure $i=\sqrt{-1}$ can be interpreted as a two blade on phase space. 
If the phase space is just two dimensional there is only one candidate
for the imaginary structure, namely the symplectic volume form
$\texttt{j}=\fett{\eta}\fett{\rho}$. That the $i$ in the Moyal
product has to be an unit area bivector can be seen from the
integral representation of the Moyal product \cite{Baker}:
\begin{equation}
(f\starM g)(q,p)=\frac{1}{\pi^2\hbar^2}\int dq'dq''dp'dp''f(q',p')
g(q'',p'')\exp\left(\frac{2}{\hbar}2iA_\triangle(\vec{x},
\vec{x}\,',\vec{x}\,'')\right),\label{starMint}
\end{equation}
where $A_\triangle(\vec{x},\vec{x}\,',\vec{x}\,'')$ is the area of
the triangle spanned by the vectors $\vec{x}=(q,p)^T$,
$\vec{x}\,'=(q',p')^T$ and $\vec{x}\,''=(q'',p'')^T$. So the $i$
plays here the role of the unit area bivector in phase space.
The two dimensional Moyal product can then be written with the
gradient $\nabla_x=\fett{\eta}\partial_q +\fett{\rho}\partial_p$
as:
\begin{equation}
f\starM g
=f\,e_{\starP}^{\frac{\hbar}{2}\lvec{\nabla}_x\vec{\nabla}_x}\,g
=f\,e_{\starP}^{\frac{\hbar}{2}\texttt{j}\left(\lvec{\nabla}_x\cdot
\widetilde{\vec{\nabla}}_x\right)}\,g
=f\,e_{\starP}^{\frac{\hbar}{2}\texttt{j}(\lvec{\partial}_q
\vec{\partial}_p-\lvec{\partial}_p\vec{\partial}_q)}\,g,
\end{equation}
so that the correspondence principle has the form
\begin{equation}
\lim_{\hbar\rightarrow
0}\frac{-1}{\hbar}\texttt{j}\starP\left[f,g\right]_{\starM}
=\{f,g\}_{PB}.
\end{equation}
One should also note the similarity to the fermionic star product of
two vectors $\fett{a}=a_1\fett{\eta}+a_2 \fett{\rho}$
and $\fett{b}=b_1\fett{\eta}+b_2 \fett{\rho}$:
\begin{equation}
\fett{a}\starP\fett{b}=|\fett{a}||\fett{b}|e_{\starP}^{
2\mbox{\footnotesize\boldmath$\eta$}
\mbox{\footnotesize\boldmath$\rho$}A_\triangle(
\mbox{\footnotesize\boldmath$a$},
\mbox{\footnotesize\boldmath$b$})},
\end{equation}
where $A_\triangle( \mbox{\boldmath$a$},\mbox{\boldmath$b$})$ is
the volume of the triangle spanned by the vectors $\fett{a}$ and
$\fett{b}$.

\section{Spacetime algebra and Dirac theory}
\setcounter{equation}{0}
Just as it is possible to describe geometric algebra as a
fermionic deformed superanalysis it is also possible to describe
spacetime algebra in this context. The basis vectors of space-time
are the Grassmann elements $\gamma_0,\ \gamma_1,\ \gamma_2$ and
$\gamma_3$, which fulfill
\begin{equation}
\gamma_{\mu}\cdot\gamma_{\nu}=\frac{1}{2}\left(\gamma_{\mu}\starP
\gamma_{\nu}+\gamma_{\nu}\starP\gamma_{\mu}\right)=g_{\mu\nu},
\end{equation}
where we choose here $g_{\mu\nu}=\mathrm{diag}(1,-1,-1,-1)$. The
corresponding Clifford star product in space-time is
\begin{equation}
F\starP G=F\,\exp\left[g_{\mu\nu}\frac{\lvec{\partial}}
{\partial\gamma_{\mu}}\frac{\vec{\partial}}{\partial\gamma_{\nu}}
\right]\, G.\label{STApauli}
\end{equation}
A general supernumber in space-time has the form
\begin{equation}
A=a_0+a^{\mu}\gamma_{\mu}+a^{\mu\nu}\gamma_{\mu}\gamma_{\nu}
+a^{\mu\nu\rho}\gamma_{\mu}\gamma_{\nu}\gamma_{\rho}+a_4 I_4,
\end{equation}
where $I_4=\gamma_0\gamma_1\gamma_2\gamma_3$ and only linear
independent terms should appear. With the four dimensional
pseudoscalar $I_4$ and the Clifford star product (\ref{STApauli})
it is possible to construct analogous to (\ref{obendual}) the dual
basis $\gamma^{\mu}$, which gives $\gamma^0=\gamma_0$ and
$\gamma^i=-\gamma_i$. Furthermore one can define in analogy to 
the three dimensional case a trace: 
\begin{equation}
\mathrm{Tr}(F)=4\int d\gamma_3d\gamma_2d\gamma_1d\gamma_0 \star F.
\end{equation}
The Berezin integral acts hereby again like a projector on the scalar 
part of $F$. The definition of the trace by projecting on the scalar
part was already given in \cite{Hamilton} and it was also stated
that the use of geometric algebra greatly simplifys all the trace 
calculations usually done in the matrix formalism. An expicit expression
for the trace can now in the formalism of deformed superanalysis be 
given by the Berezin integral.  

The question is now how a spacetime vector
$x=x^{\mu}\gamma_{\mu}$ is related to its space vector part
$\fett{x}=x^i\fett{\sigma}_i$. In the $\gamma_0$-system this can
be seen by a space-time split which amounts to star-multiplying by
$\gamma_0$:
\begin{equation}
x\starP\gamma_0=x\cdot\gamma_0+x\gamma_0=t+\fett{x}.
\end{equation}
One should note that $\fett{x}=x\gamma_0=x^1\gamma_1\gamma_0
+x^2\gamma_2\gamma_0+x^3\gamma_3\gamma_0$ is a spacetime
bivector, but on the other hand it is also a space vector because 
the two-blades
$\gamma_i\gamma_0$ behave like $\fett{\sigma}_i$:
\begin{eqnarray}
\fett{\sigma}_i\cdot\fett{\sigma}_j&=&\frac{1}{2}\left(\gamma_i\gamma_0
\starP\gamma_j\gamma_0+\gamma_j\gamma_0\starP\gamma_i\gamma_0\right)
=\delta_{ij},\nonumber\\
I_3&=&\fett{\sigma}_1\starP\fett{\sigma}_2\starP\fett{\sigma}_3=
\gamma_1\gamma_0\starP\gamma_2\gamma_0\starP\gamma_3\gamma_0=\gamma_0
\gamma_1\gamma_2\gamma_3=I_4,\nonumber\\
\fett{\sigma}_i\fett{\sigma}_j&=&\frac{1}{2}\left(\gamma_i\gamma_0
\starP\gamma_j\gamma_0-\gamma_j\gamma_0\starP\gamma_i\gamma_0\right)\nonumber\\
&=&\gamma_j\gamma_i=I_4\starP
\gamma_k\gamma_0=I_3\starP\sigma_k\qquad\mathrm{for\ cyclic}\,\,
i,j,k.\label{siggammarel}
\end{eqnarray}
Where the four dimensional star product (\ref{STApauli}) and the
three dimensional star product (\ref{starPDef2}) is used in
(\ref{siggammarel}) should be clear from the context.  The square
of the position four vector is
$x^{2\starP}=t^2-\fett{x}^{2\starP}$.

If a particle is moving in the $\gamma_0$-system along $x(\tau)$,
where $\tau$ is the proper time, the proper velocity is given by
$u(\tau)=\frac{d}{d\tau}x(\tau)$, with $u^{2\starP}=1$. For the
space-time split of the proper velocity one obtains:
\begin{equation}
u\starP\gamma_0=u\cdot\gamma_0+u\gamma_0=\frac{d}{d\tau}\left(
x(\tau)\starP\gamma_0\right)=\frac{d}{d\tau}\left(t+\fett{x}\right)
=\frac{dt}{d\tau}+\frac{d\fett{x}}{dt}\frac{dt}{d\tau}.
\end{equation}
Comparing the scalar and the bivector part leads to
\begin{equation}
u_0=u\cdot\gamma_0=\frac{dt}{d\tau}\qquad\mathrm{and}\qquad
\fett{u}=\frac{d\fett{x}}{dt}=\frac{d\fett{x}}{d\tau}
\frac{d\tau}{dt}=\frac{u\gamma_0}{u\cdot\gamma_0}
\end{equation}
and with $1=u^{2\starP}=u_0^2(1-\fett{u}^{2\starP})$ one gets 
\cite{Hestenes2}
\begin{equation}
u_0=u\cdot\gamma_0=\frac{1}{\sqrt{1-\fett{u}^{2\starP}}}=\gamma.
\end{equation}

It is now also possible to specify a Lorentz transformation from a
coordinate system $\gamma_{\mu}$ to an in $\gamma_1$-direction
moving coordinate system $\gamma_{\mu}'$. For the coefficients
this transformation is given by $t=\gamma(t'+\beta {x'}^1)$,
$x^1=\gamma({x'}^1+\beta t')$, $x^2={x'}^2$, and $x^3={x'}^3$. The
condition $x=x^{\mu}\gamma_{\mu}={x'}^{\mu}\gamma_{\mu}'$ leads
then to
\begin{equation}
\gamma_0'=\gamma(\gamma_0+\beta\gamma_1)\qquad\mathrm{and}\qquad
\gamma_1'=\gamma(\gamma_1+\beta\gamma_0).
\end{equation}
Introducing the angle $\alpha$ so that $\beta=\tanh(\alpha)$ this
can be written as
\begin{equation}
\gamma_0'=\cosh(\alpha)\gamma_0+\sinh(\alpha)\gamma_3
=e_{\starP}^{\alpha\gamma_1\gamma_0}\starP\gamma_0\qquad\mathrm{and}\qquad
\gamma_1'=\cosh(\alpha)\gamma_1+\sinh(\alpha)\gamma_0
=e_{\starP}^{\alpha\gamma_1\gamma_0}\starP\gamma_1
\label{Lotra2}
\end{equation}
or with $L_1=e_{\starP}^{\alpha\gamma_3\gamma_0/2}$ as
$\gamma_{\mu}'=L_1\starP\gamma_{\mu}\starP\overline{L_1}$. In
general the generators of a passiv Lorentz transformation can be
calculated with
\begin{equation}
\sigma_{\mu\nu}
=\frac{I_4}{2}\starP\left[\gamma_{\mu},\gamma_{\nu}\right]_{\starP},
\end{equation}
so that the generators for the boosts and the rotations are
\begin{equation}
K_i=\frac{1}{2}\sigma_{0i}\qquad\mathrm{and}\qquad
S_i=\frac{1}{2}\sum_{j<k}\varepsilon_{ijk}\sigma_{jk}.
\end{equation}
These generators satisfy
\begin{equation}
\left[S_i,S_j\right]_{\starP}=I_4\starP\varepsilon^{ijk}S_k,\qquad
\left[S_i,K_j\right]_{\starP}=I_4\starP\varepsilon^{ijk}K_k,\quad\mathrm{and}\quad
\left[K_i,K_j\right]_{\starP}=-I_4\starP\varepsilon^{ijk}S_k
\end{equation}
and a passive Lorentz transformation is given by
\begin{equation}
\gamma_{\mu}'=\Lambda_{\mu}^{\,\nu}\gamma_{\nu}=
e_{\starP}^{\frac{1}{4}I_4\starP\sigma_{\mu\nu}\omega^{\mu\nu}}
\starP\gamma_{\mu}\starP
e_{\starP}^{-\frac{1}{4}I_4\starP\sigma_{\mu\nu}\omega^{\mu\nu}},
\end{equation}
which is a generalization of (\ref{Lotra2}).

The Dirac equation can then be written down immediately as \cite{Deform5}
\begin{equation}
(p\mp m)\starMP\pi_{\pm m}^{(MC)}(p)=0,
\end{equation}
where no slash notation is needed, because one naturally has
$p=p^{\mu}\gamma_{\mu}$. The Wigner function $\pi_{\pm
m}^{(MC)}(p)$ for the Dirac equation is the functional analog of
the well known energy projector of Dirac theory:
\begin{equation}
\pi_{\pm m}^{(MC)}(p)=\frac{\pm p+m}{2m}.
\end{equation}
Besides the energy one also has the spin as an observable, which
is here given by
\begin{equation}
S_s=\frac{\hbar}{2}\gamma_5\starP s,
\end{equation}
where $s=s^{\mu}\gamma_{\mu}$ is a vector which fulfills
$s^{2\starP}=-1$ and $s\cdot p=0$. $\gamma_5$ is here
$\gamma_5=iI_4$.
With $S_s\starP S_s=\left(\frac{\hbar}{2}\right)^2$ and
$\left[S_s,p\mp m\right]_{\starP}=0$ one sees that the spin Wigner
function is given by the functional analog of the spin projector
in Dirac theory
\begin{equation}
\pi_{\pm s}^{(C)}(s)=\frac{1}{2}\pm\frac{1}{\hbar}S_s
\end{equation}
and fulfills $S_s\starP\pi_{\pm s}^{(C)}(s)=\pm\frac{\hbar}{2}
\pi_{\pm s}^{(C)}(s)$. The total Wigner function is then the
Clifford star product of the two single Wigner functions.

\section{Conclusions}
\setcounter{equation}{0}

\qquad There are two formal and conceptual barriers that separate
quantum theory from classical theory. The first barrier is that
classical theory is described on the phase space while quantum
theory is described on the Hilbert space. This conceptual barrier
is overcome by the program of deformation quantization that
describes quantum theory on the phase space. The second barrier is
that one uses in classical mechanics the
Gibbs-Heaviside formalism, which can not take spin into account.
In quantum theory where spin is a physical observable it is
described in the nonrelativistic case by the Feynman trick, which
substitutes $\vec{p}$ by $\vec{p}\cdot\vec{\sigma}$ and in the
relativistic case it is introduced by writing
$p_{\mu}\gamma^{\mu}$. Both notations clearly indicate that the
$\sigma_i$ and the $\gamma_{\mu}$ are basis vectors, but this is
obscured by representing them by matrices. The work of Hestenes
has clarified this point by formulating classical and quantum
theory in the same formalism of geometric algebra. The astonishing
thing is now that also this second barrier can be overcome in terms
of the star product formalism, so that classical and quantum theory
can be unified on a formal level. Both can be described by the
formalism of deformed superanalysis, where classical mechanics is
a half deformed formalism, that means the deformation only takes
place in the Grassmann sector of superanalysis, while quantum
mechanics leads to a totaly deformed formalism, where also the
product of the scalar coefficients is deformed. This shows at 
least on a formal level in which way quantum theory is a more 
fundamental theory compared to the classical theory.

The star product formalism has also advantages in the context of
geometric calculus, because it gives an explicit expression for
the geometric product. Geometric algebra in the way Hestenes
constructed it, is formulated with respect to the scalar and the
wedge product, which represent the lowest and the highest order
terms of the geometric product. All other terms of the geometric
product are then formulated with the help of these two products.
This approach is very practical, especially if one has only terms
that are at most bivectors. But in the general case the highest
and the lowest terms of an expansion have on a formal level the
same status as all other terms. The star product gives now all
these terms of different grade as terms of an expansion, that can
be calculated in a straight forward fashion.


\end{document}